\documentclass{ws-procs9x6}

\begin{document}

\title{How Well Does AdS/QCD Describe QCD?}

\author{J. ERLICH}

\address{Physics Department, College of William and Mary\\
Williamsburg, VA, 23185, USA \\
E-mail: erlich@physics.wm.edu}

\begin{abstract}
AdS/QCD is an extra-dimensional approach to modeling hadronic physics, 
motivated by the AdS/CFT correspondence in string theory.  
AdS/QCD models are often
more accurate than would have been expected at energies below a few GeV.  
We address the question of why these models are so successful, and 
respond to
some of the criticisms that have been waged against these models.
\end{abstract}


\bodymatter

\section{What is AdS/QCD?}\label{sec:AdSQCD}
The AdS/CFT correspondence \cite{Maldacena:1997re} provides a powerful tool for
extra-dimensional model-building.  Qualitative features of
electroweak symmetry breaking models with warped extra dimensions 
 can
often be predicted by analogy with the AdS/CFT correspondence
\cite{ArkaniHamed:2000ds}.
Even quantitatively,
simple extra-dimensional models of QCD motivated by the AdS/CFT
correspondence have proven successful at reproducing low-energy
hadronic data like meson masses, decay constants, and coefficients of the
chiral Lagrangian \cite{deTeramond:2005su,Erlich:2005qh,DaRold:2005zs}.  
These models fall into two classes:
top-down models based on brane constructions in string theory, and
bottom-up models which are more phenomenological.  Both top-down and 
bottom-up models in this framework are referred to as AdS/QCD models 
or holographic QCD.  

AdS/QCD models are related to a number of earlier ideas which are 
useful for
understanding properties of QCD at low energies: chiral symmetry
breaking, hidden local symmetry \cite{Bando:1984ej}, vector meson
dominance \cite{Sakurai:1969ss}, large $N$
\cite{'tHooft:1973jz}, the Weinberg and Shifman-Vainshtein-Zakharov (SVZ)
sum rules \cite{Weinberg:1967kj,Shifman:1978bx}, Skyrmions
\cite{Skyrme:1962vh}, and matching of the low energy and high energy regimes
\cite{Migdal}.  
Aside from the pattern of
chiral symmetry breaking, 
which is an input of some AdS/QCD models, aspects of all of
the features listed above are a natural consequence of the
extra-dimensional nature of these models.  Hidden local symmetry is
related to the gauging of the global symmetry as suggested by the AdS/CFT
correspondence \cite{Son:2003et}; 
large $N$ is
required by the classical limit of the AdS/CFT correspondence 
\cite{Maldacena:1997re} and is reflected in poles in correlation functions
at real values of squared momenta;
dominance of the lightest rho meson in hadronic couplings is due to
oscillations of the heavier Kaluza-Klein modes in the extra dimensions
\cite{Son:2003et,Hong:2005np}; 
sum rules become exact relations for sums over Kaluza-Klein modes 
\cite{Erlich:2005qh,DaRold:2005zs,Hirn:2005nr,Hirn:2007bb}; 
solitonic solutions
describing baryons in these models are closely related to
the Skyrme model \cite{Sakai:2004cn,Nawa:2006gv,Pomarol:2009hp}; 
resonance masses
and decay constants conspire to reproduce the high energy behavior of
correlators \cite{Shifman:2005zn,EmergingHolo}.

The AdS/CFT correspondence relates certain strongly coupled large-$N$ gauge
theories to weakly coupled theories containing gravity
\cite{Maldacena:1997re}.  The prototypical
example is ${\cal N}$=4 supersymmetric SU$(N)$ gauge
theory in the limit $N\rightarrow\infty$ and 't Hooft coupling $g^2N\gg 1$,
which is identified with Type IIB supergravity in an AdS$_5\times$S$^5$
spacetime background with nonvanishing 5-form flux.  
The five-sphere S$^5$  
plays an important role in matching supergravity fields
to operators in the ${\cal N}$=4 gauge theory, but will
not play an important role in our discussion of QCD-like models.
Anti-de Sitter space in 4+1 dimensions, AdS$_5$, is described by the metric
\begin{equation}
ds^2=\frac{R^2}{z^2}\left(\eta_{\mu\nu}\,dx^\mu\, dx^\nu-dz^2\right),
\label{eq:AdSmetric}\end{equation}
where $\mu,\nu\in(0,1,2,3)$ and $\eta_{\mu\nu}$ is the Minkowski tensor in
3+1 dimensions (4D).  The coordinate $z$ is sometimes referred to as
the radial coordinate.  The region $z>0$ covers half (a Poincar\'e patch)
of the
full AdS spacetime, which is enough for our purposes.  
Although Anti-de Sitter space is a maximally symmetric spacetime,
the radial coordinate plays a distinguished role in the
metric Eq.~(\ref{eq:AdSmetric}).
An interesting relationship between the 
radial coordinate
and light-front dynamics has been observed by Brodsky and
De Teramond (see, for example, Refs.~\cite{Brodsky:2003px,deTeramond:2008ht}).

\subsection{The top-down approach}
\begin{figure}
\psfig{file=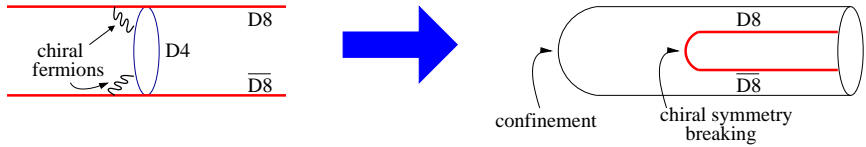,width=4.5in}
\caption{The Sakai-Sugimoto model.}
\label{fig:SS}
\end{figure}

In top-down AdS/QCD models, a brane construction in string theory is
engineered which at low energies describes a gauge theory with features
similar to QCD 
\cite{Klebanov:2000hb,Kruczenski:2003uq,Babington:2003vm,Sakai:2004cn,Antonyan:2006vw}.  
The top-down model most similar to QCD is the Sakai-Sugimoto model 
\cite{Sakai:2004cn,Sakai:2005yt}, in which
$N$ D4-branes in Type IIA string theory are wrapped on a circle, and are 
intersected by $N_f$ D8-branes and $N_f$ $\overline{{\rm D8}}$-branes,
as in Fig.~\ref{fig:SS}.
Open strings with both ends on the D4-branes describe a SU$(N)$ gauge theory
with sixteen supercharges.  However, the supersymmetry is broken if
fermions are given antiperiodic boundary conditions around the circle, 
and at low energies the system is described
by a pure SU$(N)$ gauge theory with fermions and scalar fields whose masses
are determined by the size of the circle \cite{Witten:1998zw}.  
(Unfortunately, the mass gap is comparable to the masses of the
extraneous fermions and scalars in this model.)
The D8 and $\overline{{\rm D8}}$-branes also break
the supersymmetry.  A string with one end on a D4-brane and one end on a
D8-brane ($\overline{{\rm D8}}$-brane) 
describes a left-handed (right-handed) fermion in the effective 4D theory.  

The separation of left and right-handed fermions makes chiral symmetry 
manifest in this theory, 
and makes the Sakai-Sugimoto model the AdS/CFT model of choice for 
describing QCD.
The gravitational backreaction of the 
D4-branes generates  
a horizon, which is related to confinement as the horizon creates a 
mass gap in gravitational fluctuations 
which are interpreted as glueballs \cite{Witten:1998zw,Csaki:1998qr}.  
Chiral symmetry 
breaking in the Sakai-Sugimoto model is due to the fact that the D8-branes
and $\overline{{\rm D8}}$-branes are joined together as a result of the
curved spacetime geometry, as in Fig.~\ref{fig:SS}.  
In this model the location at which the D8 and
$\overline{{\rm D8}}$-branes meet is an adjustable parameter and controls the
ratio of confining to chiral symmetry breaking scales 
\cite{Antonyan:2006pg,Aharony:2006da,Carone:2007md}.

\subsection{The bottom-up approach}
In bottom-up AdS/QCD models, we specify an extra-dimensional 
spacetime geometry and the fields that propagate in them based on the
properties of QCD which we would like to be incorporated.  Towers of
Kaluza-Klein modes are identified with
towers of radial excitations of QCD states. 
Boundary conditions on gauge fields break
the higher-dimensional gauge invariance, while the corresponding global 
symmetry
remains in the effective 3+1 dimensional theory.  For example,
an SU(2)$\times$SU(2) gauge theory in 4+1 dimensions (5D) can represent the
approximate
chiral symmetry of QCD.  A set of scalar fields transforming in the fundamental
representation of both SU(2) gauge groups can spontaneously
break the symmetry to a diagonal
subgroup which may be identified with isospin.  This is the basic scenario
in some basic 
bottom-up models \cite{Erlich:2005qh,DaRold:2005zs}.  Rather than use scalar
fields to break the chiral symmetry, modified spacetime geometry and
boundary conditions can do
the same \cite{Hirn:2005nr,Hirn:2005vk}.  
A more general
approach includes the scalar fields with modified boundary conditions
\cite{Erlich:2008gp}.

For simplicity the spacetime geometry in bottom-up models is often chosen
to be a slice of AdS$_5$ with metric given by Eq.~(\ref{eq:AdSmetric}) between
$z=\epsilon$ and $z=z_m$, where $\epsilon/z_m\ll1$.  Such models are
called hard-wall models because of the sharp boundary at $z=z_m$ 
\cite{Polchinski:2000uf,Karch:2002xe,Hong:2004sa,Erlich:2005qh,DaRold:2005zs}.  The AdS$_5$
spacetime has an SO(2,4) isometry which is broken only by
the boundaries of the spacetime.  The consequence is that the effective
3+1 dimensional theory has an SO(2,4) symmetry at high energies which
is identified with the conformal invariance of QCD at high energies.
The conformal symmetry ensures that correlation functions of operators
as derived in these models will have the form required by asymptotic freedom.
However, other aspects of these models are not expected to be valid at
energies above a few GeV where stringy effects become important in QCD,
so the matching of certain predictions of AdS/QCD models to the UV
is somewhat {\em ad hoc}.  

In summary, a basic bottom-up model is described by an SU(2)$\times$SU(2)
gauge theory in a slice of AdS$_5$ with bifundamental scalar field $X$
and action \begin{eqnarray}
S&=&\int d^4x\int_\epsilon^{z_m}dz 
\sqrt{|g|} \left[-1/(2g_5^2)\ {\rm Tr}
\left(L_{MN}L^{MN}+R_{MN}R^{MN}\right) \right. \\
&&\left.+{\rm Tr} \left(|D_M X|^2-m_X^2|X|^2\right)\right],
\end{eqnarray}
where Lorentz indices $M$ and $N$ run from 0 to 4, and 
$L_{MN}$ and $R_{MN}$ are
field stregths of the two sets of SU(2) gauge fields.  
The fields $X_{ij}$, $i,j\in(1,2)$, 
have the quantum numbers of the scalar
quark bilinear $\overline{q}_{Li}q_{Rj}$, where the subscripts $L$ and
$R$ refer to left and right-handed chirality, respectively, and $q_1$
and $q_2$ are the up and down quark fields, respectively.  According to 
the AdS/CFT
correspondence we would identify the mass $m_X$ of the field dual to an 
operator of scaling dimension $\Delta=3$, via 
\cite{Witten:1998qj,Gubser:1998bc}
\begin{equation}
m_X^2 R^2=\Delta(\Delta-4)=-3, \label{eq:mX}\end{equation}
where $R$ is the AdS curvature scale of 
Eq.~(\ref{eq:AdSmetric}).  
We may make this choice for definiteness, although the
AdS/CFT correspondence is not strictly valid for this model and we 
would be
ignoring renormalization 
effects which modify operator scaling dimensions away from
the high energy regime.

The solutions to the classical equations of motion with vanishing gauge fields
and $X$ field of the form $X_0(z)$ are \begin{equation}
X_0(z)=m_q \,z^{4-\Delta}+\frac{\sigma}{4\Delta-8} \,z^\Delta. 
\label{eq:Xbackgnd}\end{equation}
For $2<\Delta<4$, the term proportional to the parameter $m_q$
is non-normalizable in the sense that the contribution to the action from
the integral over
$z$ diverges if the UV cutoff length $\epsilon\rightarrow0$.  On the other
hand, the term proportional to $\sigma$ is normalizable.  
We will allow both normalizable and non-normalizable backgrounds for the field
$X$.  The factor of $1/(4\Delta-8)$ in Eq.~(\ref{eq:Xbackgnd}) is motivated
by the AdS/CFT identification of $\sigma$ as the (isospin-preserving)
chiral condensate $\left<\overline{u}_Lu_R\right>=
\left<\overline{d}_Ld_R\right>$ \cite{Klebanov:1999tb}.

The predictions
of the model are  insensitive to $\epsilon$ as long as $\epsilon/z_m\ll1$.
Hence, the free parameters in the model are $m_X$, $g_5$, $m_q$, $\sigma$,
and $z_{IR}$.  If we fix $m_X^2=-3/R^2$ then there are four parameters
in addition to the choice of boundary conditions on the gauge fields.
We can also fix $g_5$ by extending AdS/QCD predictions to the UV and matching
with perturbative QCD, leaving only three model parameters \cite{Erlich:2005qh,DaRold:2005zs}.
However, this model is especially simplistic.  For example, 
there could be a nontrivial potential for $X$, as in 
J. Kapusta's talk \cite{Kapusta}, and the geometry does not need to
be Anti-de Sitter space \cite{Hirn:2005vk,Becciolini:2009fu}.

Meson masses and decay constants are determined by the 
effective 3+1 dimensional theory.  The Kaluza-Klein modes of the gauge fields
have the quantum numbers of vector and axial-vector mesons. The masses of
the Kaluza-Klein modes are determined by the eigenvalues of the equations
of motion together with the prescribed boundary conditions.  The decay
constants are determined by computing the mixing between the Kaluza-Klein modes
and the zero modes for the gauge fields 
\cite{Sakai:2004cn,Sakai:2005yt}.  Alternatively,
the AdS/CFT correspondence can be used to calculate the correlation function
of products of currents, and the the decay constants are the residues of
poles at the location of the Kaluza-Klein masses \cite{Erlich:2005qh,DaRold:2005zs}.
Some of the early predictions of this model, with particular values of the
parameters, appear in Table~\ref{tab:EKSS}.  Some of the coefficients
of the chiral Lagrangian are related to these observables, and were
calculated in the same model with a different choice of parameters in
Ref.~\cite{DaRold:2005zs}.
\begin{table}
\tbl{Predictions of a hard-wall model.  For more details see Refs.
\cite{Erlich:2005qh,Katz:2005ir}.}
{\begin{tabular}{|c|c|c|c|}  
\hline
Observable         & Measured                & Model \\
                   & (MeV, central values)                   & (MeV) \\ \hline
$m_\pi$            & 139.6        & 141 \\
$m_\rho$           & 775.8          & 832   \\ 
$m_{a_1}$          & 1230             & 1220  \\
$f_\pi$            & 92.4           & 84.0  \\
$F_\rho^{\,1/2}$   & 345               & 353   \\
$F_{a_1}^{\,1/2}$  & 433              & 440   \\
$g_{\rho\pi\pi}$   & 6.03                  & 5.29  \\ 
$m_{f_2}$          & 1275          & 1240 \\ \hline
\end{tabular}}
\label{tab:EKSS}
\end{table}

Numerous additional results in AdS/QCD models have
been calculated.  In a hard-wall AdS/QCD model 
including a strange quark mass parameter, predictions
for masses and decay constants of light mesons containing strange quarks
mostly agree with 
experimental central values at the few percent level \cite{KatzLatz}.  
The $q^2$ dependence of form factors and moments of
generalized parton distributions in various
AdS/QCD models are in relatively good agreement with experiment
up to a few GeV (see, for example, 
Refs.~\cite{Kwee:2007nq,Grigoryan:2007vg,Abidin:2008ku}).
From the behavior of form factors at small $q^2$, 
these models predict that mesons have a larger
size as determined by their charge distribution than by their momentum
distribution \cite{Abidin:2008ku}.

\section{How well does AdS/QCD work?}

The summary of our discussion thus far is that AdS/QCD 
models seem to be reasonably reliable for predicting observables
at below a few GeV, but tend to make poor predictions at higher energies.  
Until we better understand why certain models
work especially well, we should at best trust only those predictions which are
independent of model details such as the choice
of spacetime geometry or boundary conditions.  

\begin{figure}
\psfig{file=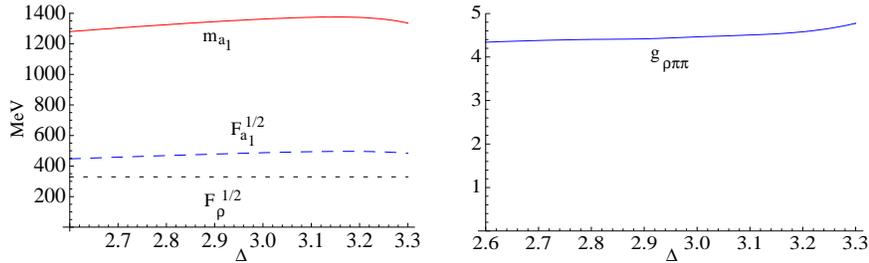,width=4.5in}
\caption{AdS/QCD predictions are not sensitive to varying $\overline{q}q$
scaling dimension 
$\Delta$ around its UV value $\Delta_{UV}=3$, from Ref.~\cite{Erlich:2008gp}.}
\label{fig:Delta}
\end{figure}

According to the AdS/CFT correspondence, finite temperature physics can
be studied by introducing a black hole into the higher-dimensional
spacetime.  From general properties of black hole horizons, a universal
prediction of the ratio of shear viscosity $\eta$ to entropy density $s$
was discovered by Kovtun, Son and Starinets \cite{Kovtun:2004de}:
$\eta/s=1/(4\pi)$. 
Other universal relations have subsequently been found, such as a relation
between
electrical conductivity to charge susceptibility 
\cite{Kovtun:2008kx}.
The zero-temperature observables quoted earlier are not universal, but we
can test their dependence on model parameters.  In the hard-wall model
defined in the previous section, allowing the effective dimension 
$\Delta$ of the operator $\overline{q}q$ to vary from its UV value 
$\Delta_{UV}=3$ is tantamount to varying the mass of the field $X$ via
Eq.~(\ref{eq:mX}).  The mass varies from the Breitenlohner-Freedman bound
\cite{Breitenlohner:1982jf} $m_X^2=-4/R^2$ for $\Delta=2$, 
to $m_X^2=0$ at $\Delta=4$.  The
dependence of some AdS/QCD predictions as a function of $\Delta$ over part
of this range is shown in Fig.~\ref{fig:Delta}, holding fixed $m_\pi$, 
$f_\pi$ and $m_\rho$ \cite{Erlich:2008gp}.   
Note
the insensitivity of hard-wall 
AdS/QCD predictions to the parameter $\Delta$.  However,
there is more sensitivity if the
boundary conditions on the gauge fields in the IR are modified 
\cite{Erlich:2008gp}.  
If the geometry is allowed to vary in a certain
class of AdS/QCD models, it seems that
the AdS geometry provides an especially good fit to data,
but certain predictions are relatively insensitive to the details of the
choice of geometry \cite{Becciolini:2009fu}.  The hard-wall model has a 
root-mean-squared error with seven observables and three parameters
(as defined in Ref.~\cite{Erlich:2005qh}) of under 10\%.
In general, AdS/QCD models
are consistent with one another at around the 20-25\% level, and
I am hopeful that we will better understand the
unreasonable success of specific models.  

\section{Challenges for AdS/QCD}
I would like to address some of the criticisms of the AdS/QCD approach
that have been raised, both in the literature and in private (and 
not-so-private) communication.  
I contend that most of these
objections are red herrings if the scope
of AdS/QCD models is properly understood, but AdS/QCD still faces some
phenomenological challenges.

\begin{itemize}
\item {\em AdS/QCD is not AdS/CFT.}
\end{itemize}

The ${\cal N}$=4 Yang-Mills theory which serves as the prototypical example
of the AdS/CFT correspondence is supersymmetric, conformal, and has a large
number of colors and large 't Hooft coupling for calculability.  This is all
quite
different from QCD, so why should we rely on the AdS/CFT correspondence to
model QCD?

First of all, it is important to note that the existence of gravity duals 
is not tied to conformal invariance or
supersymmetry.  The Sakai-Sugimoto model is an example without either 
feature.  
The large-$N$ limit is required for a classical description 
with decoupled stringy
physics, and the narrow resonances predicted by large $N$ are indeed features
of classical AdS/QCD calculations.  It would be interesting to pursue
$1/N$ corrections from quantum corrections in the 5D models, and as far as 
I know little
work has been done in this regard (although there are close analogies to
5D models of electroweak symmetry breaking for which quantum corrections
have been studied).  For example, in AdS/QCD models  
the $\rho\pi\pi$ coupling has been calculated
and compared with the decay rate of the rho meson.  Such couplings 
can be used to self-consistently modify corresponding spectral functions.

The analogy with the
AdS/CFT correspondence is useful for inferring qualitative features
of AdS/QCD models, but these models
stand on their own as 5D models independent of the 
AdS/CFT correspondence.

\begin{itemize}
\item {\em The spectrum is not Regge-like.}
\end{itemize}

Misha Shifman noted that hard-wall AdS/QCD models have spectra which
depend on radial excitation number $n$ like $m_n^2\sim n^2$ as opposed to the 
Regge-like behavior $m_n^2\sim n$ \cite{Shifman:2005zn}.  
The soft-wall model \cite{Karch:2006pv} obtains the correct Regge-like spectrum
by introducing a dilaton background which modifies the equations of motion
appropriately.
At low energies the predictions of the soft-wall and hard-wall models are 
comparable.

\begin{itemize}
\item {\em The model is matched to the UV, where QCD is weakly coupled.}
\end{itemize}

In three places the hard-wall model described above
is matched to the UV: in the choice of AdS$_5$ geometry, in specifying the
5D gauge coupling $g_5$ by comparison with UV current correlators, and in
assigning the squared mass $m_X^2=-3/R^2$ to the scalar field by analogy 
with the 
AdS/CFT correspondence and the UV dimension of the operator $\overline{q}q$.  
Matching to the UV  
was done for definiteness, but one can allow parameters
to vary, as in Ref.~\cite{Erlich:2008gp}.  The conclusion is that
matching to experiment prefers $g_5$ fixed to around the value obtained by
matching to the UV, while there is less sensitivity to variation of $m_X$.

On a related note, 
Misha Shifman has pointed out that AdS/QCD predictions for
correlators of isospin currents agree with Migdal's
Pad\'e approximation
for the UV behavior of current correlators \cite{Shifman:2005zn}.
The interesting thing about this approach is that one can systematically 
observe
the radial direction of Anti-de Sitter space open up as the number of 
poles included in the Pad\'e approximation increases \cite{EmergingHolo}.

\begin{itemize}
\item {\em Identification of QCD parameters depends on $N$.}
\end{itemize}

The pattern of chiral symmetry breaking is built into
the AdS/QCD models we have been discussing.  
If the AdS/CFT correspondence were strictly
valid for this model, then $m_q$ and $\sigma$ of Eq.~(\ref{eq:Xbackgnd}) would
play the role of the source and expectation value of the operator 
$\overline{q}_L q_R$ dual to the scalar field $X$, up to an $N$-dependent 
rescaling \cite{Cherman:2008eh}.  The $N$ dependence is not included in 
Eq.~(\ref{eq:Xbackgnd}), so we should allow the possibility of rescaling 
$m_q$ and $\sigma$ appropriately.
On the other hand, if we treat $m_q$ and $\sigma$ simply as model parameters,
then the interpretation of those parameters
as quark mass and chiral condensate is understood to be imprecise from
the outset.

\begin{itemize}
\item {\em AdS/QCD has incorrect high energy behavior.}
\end{itemize}

Even though Anti-de Sitter space guarantees correct conformal behavior at
high energies and the soft-wall model produces Regge-like spectra, 
there are many things AdS/QCD continues to get wrong at high energies.
The density of states does not agree with the spectrum expected from string
theory, and high-energy scattering amplitudes are not correctly predicted
\cite{Hofman:2008ar,Csaki:2008dt,pire}.  

We are forced at the moment
to understand these models as low-energy descriptions valid only below some 
cutoff scale which may differ between models but is generally a few GeV.
It is still valuable to try to match certain aspects of these models to the
UV,  but at the same time we have to recognize the limitations of these
models at high energies.  Perhaps we can reconcile AdS/QCD predictions with
high-energy scattering by including progressively more 5D fields
into the model
as the energy scale of interest increases, although so doing would introduce 
more model parameters and a decrease in predictivity.

\begin{itemize}
\item {\em Important operators are not accounted for.}
\end{itemize}

The glueball condensate is important for certain observables but
is not included in some of these models.
Simple AdS/QCD models like the hard-wall and soft-wall models also do not
include fields which would be dual to the operators $\overline{q}
[\gamma^i,\gamma^0]T^a q$, which have significant overlap between vacuum
and one-rho meson states \cite{Glozman}.  
The basic response to such objections is that there
is a natural way to include fields into the 5D theory which would describe
corresponding operators and condensates in the effective 4D theory.  
The glueball condensate is related to the dilaton profile.  New 5D fields can
be included with quantum numbers conjugate to the operators in question, 
and one should study the
effects of those fields.  Any 5D field charged under the gauged isospin
symmetry will lead to new couplings involving the rho meson which can be
compared with experiment.
At higher energies, more 5D fields are expected to be required
in order to describe higher-dimension 4D operators that become important at
those scales.

\begin{itemize}
\item {\em AdS/QCD is an uncontrolled expansion.}
\end{itemize}

If we understand AdS/QCD as a low-energy effective theory, then we should
expect quantum corrections to generate 
higher-dimension operators in the 5D action suppressed by some scale.
In AdS/QCD the relevant scale is the confining scale (or the chiral 
symmetry breaking scale, which is comparable).
However, we are interested in physics above that scale.
This is an honest
difficulty for AdS/QCD models, and deserves further exploration.  

Higher-dimension operators are generated by integrating out fields
above the scale of interest.  
In the large-$N$ limit with large 't Hooft coupling, high-spin 
states decouple from the theory.\footnote{I thank Kaustubh Agashe for discussion on this point.}
Perhaps the effectiveness of the large-$N$ expansion in this limit 
is also related to
the effectiveness of the AdS/QCD approach at low energies, summing only over
``supergravity states'' with spin less than or equal to two.

\section{Conclusions}
In summary, AdS/QCD models are extra-dimensional models of QCD resonances,
and are generally accurate below a few GeV at the 10-25\% level depending
on the details of the model.    
The benefit of an extra-dimensional approach is that several
features of QCD are immediate consequences of extra dimensions.  
QCD
sum rules, vector meson dominance, and 
hidden local symmetry 
are all natural features of extra-dimensional models.
A number of criticisms 
have been raised which challenge the AdS/QCD approach, based on the 
comparison of AdS/QCD models with the AdS/CFT correspondence in string theory
and based on 
flawed phenomenology of these models, especially at high energies.  Most of
these objections disappear if we understand AdS/QCD as a class of 
effective low-energy models independent of the AdS/CFT correspondence.

It is an honor and a pleasure to have been part of this wonderful conference
celebrating the vast achievements of Misha Shifman, who shows no sign of 
slowing down.
Misha has played an important role in identifying challenges for
AdS/QCD 
and has influenced the development of this field over the past few
years. Thanks, and happy birthday Misha.


\begin{thebibliography}{99}
\bibitem{Maldacena:1997re}
  J.~M.~Maldacena,
  Adv.\ Theor.\ Math.\ Phys.\  {\bf 2}, 231 (1998)
  [Int.\ J.\ Theor.\ Phys.\  {\bf 38}, 1113 (1999)]
  [arXiv:hep-th/9711200].

\bibitem{ArkaniHamed:2000ds}
 N.~Arkani-Hamed, M.~Porrati and L.~Randall,
  JHEP {\bf 0108}, 017 (2001)
  [arXiv:hep-th/0012148].

\bibitem{deTeramond:2005su}
 G.~F.~de Teramond and S.~J.~Brodsky,
  Phys.\ Rev.\ Lett.\  {\bf 94}, 201601 (2005)
  [arXiv:hep-th/0501022].


\bibitem{Erlich:2005qh}
  J.~Erlich, E.~Katz, D.~T.~Son and M.~A.~Stephanov,
  Phys.\ Rev.\ Lett.\  {\bf 95}, 261602 (2005)
  [arXiv:hep-ph/0501128].

\bibitem{DaRold:2005zs}
 L.~Da Rold and A.~Pomarol,
  Nucl.\ Phys.\  B {\bf 721}, 79 (2005)
  [arXiv:hep-ph/0501218].

\bibitem{Bando:1984ej}
  M.~Bando, T.~Kugo, S.~Uehara, K.~Yamawaki and T.~Yanagida,
  Phys.\ Rev.\ Lett.\  {\bf 54}, 1215 (1985).

\bibitem{Sakurai:1969ss}
  J.~J.~Sakurai,
  Phys.\ Rev.\ Lett.\  {\bf 22}, 981 (1969).

\bibitem{'tHooft:1973jz}
 G.~'t Hooft,
  Nucl.\ Phys.\  B {\bf 72}, 461 (1974).

\bibitem{Weinberg:1967kj}
  S.~Weinberg,
  Phys.\ Rev.\ Lett.\  {\bf 18}, 507 (1967).

\bibitem{Shifman:1978bx}
  M.~A.~Shifman, A.~I.~Vainshtein and V.~I.~Zakharov,
  Nucl.\ Phys.\  B {\bf 147}, 385 (1979).

\bibitem{Skyrme:1962vh}
  T.~H.~R.~Skyrme,
  Nucl.\ Phys.\  {\bf 31}, 556 (1962).

\bibitem{Migdal}
A.~A.~Migdal,
  Annals Phys.\  {\bf 109}, 365 (1977).

\bibitem{Son:2003et}
 D.~T.~Son and M.~A.~Stephanov,
  Phys.\ Rev.\  D {\bf 69}, 065020 (2004)
  [arXiv:hep-ph/0304182].

\bibitem{Hong:2005np}
 S.~Hong, S.~Yoon and M.~J.~Strassler,
  arXiv:hep-ph/0501197.

\bibitem{Hirn:2005nr}
  J.~Hirn and V.~Sanz,
  JHEP {\bf 0512}, 030 (2005)
  [arXiv:hep-ph/0507049].

\bibitem{Hirn:2007bb}
J.~Hirn and V.~Sanz,
  Phys.\ Rev.\  D {\bf 76}, 044022 (2007)
  [arXiv:hep-ph/0702005].

\bibitem{Sakai:2004cn}
 T.~Sakai and S.~Sugimoto,
  Prog.\ Theor.\ Phys.\  {\bf 113}, 843 (2005)
  [arXiv:hep-th/0412141].

\bibitem{Nawa:2006gv}
  K.~Nawa, H.~Suganuma and T.~Kojo,
  Phys.\ Rev.\  D {\bf 75}, 086003 (2007)
  [arXiv:hep-th/0612187].

\bibitem{Pomarol:2009hp}
 A.~Pomarol and A.~Wulzer,
  arXiv:0904.2272 [hep-ph].

\bibitem{Shifman:2005zn}
  M.~Shifman,
  arXiv:hep-ph/0507246.

\bibitem{EmergingHolo}
J.~Erlich, G.~D.~Kribs and I.~Low,
  Phys.\ Rev.\  D {\bf 73}, 096001 (2006)
  [arXiv:hep-th/0602110].



\bibitem{Brodsky:2003px}
S.~J.~Brodsky and G.~F.~de Teramond,
  Phys.\ Lett.\  B {\bf 582}, 211 (2004)
  [arXiv:hep-th/0310227].


\bibitem{deTeramond:2008ht}
G.~F.~de Teramond and S.~J.~Brodsky,
  Phys.\ Rev.\ Lett.\  {\bf 102}, 081601 (2009)
  [arXiv:0809.4899 [hep-ph]].


\bibitem{Klebanov:2000hb}
 I.~R.~Klebanov and M.~J.~Strassler,
  JHEP {\bf 0008}, 052 (2000)
  [arXiv:hep-th/0007191].

\bibitem{Kruczenski:2003uq}
 M.~Kruczenski, D.~Mateos, R.~C.~Myers and D.~J.~Winters,
  JHEP {\bf 0405}, 041 (2004)
  [arXiv:hep-th/0311270].


\bibitem{Babington:2003vm}
J.~Babington, J.~Erdmenger, N.~J.~Evans, Z.~Guralnik and I.~Kirsch,
  Phys.\ Rev.\  D {\bf 69}, 066007 (2004)
  [arXiv:hep-th/0306018].



\bibitem{Antonyan:2006vw}
 E.~Antonyan, J.~A.~Harvey, S.~Jensen and D.~Kutasov,
  arXiv:hep-th/0604017.


\bibitem{Sakai:2005yt}
T.~Sakai and S.~Sugimoto,
  Prog.\ Theor.\ Phys.\  {\bf 114}, 1083 (2005)
  [arXiv:hep-th/0507073].


\bibitem{Witten:1998zw}
E.~Witten,
  Adv.\ Theor.\ Math.\ Phys.\  {\bf 2}, 505 (1998)
  [arXiv:hep-th/9803131].


\bibitem{Csaki:1998qr}
 C.~Csaki, H.~Ooguri, Y.~Oz and J.~Terning,
  JHEP {\bf 9901}, 017 (1999)
  [arXiv:hep-th/9806021].


\bibitem{Antonyan:2006pg}
E.~Antonyan, J.~A.~Harvey and D.~Kutasov,
  Nucl.\ Phys.\  B {\bf 784}, 1 (2007)
  [arXiv:hep-th/0608177].


\bibitem{Aharony:2006da}
O.~Aharony, J.~Sonnenschein and S.~Yankielowicz,
  Annals Phys.\  {\bf 322}, 1420 (2007)
  [arXiv:hep-th/0604161].


\bibitem{Carone:2007md}
C.~D.~Carone, J.~Erlich and M.~Sher,
  Phys.\ Rev.\  D {\bf 76}, 015015 (2007)
  [arXiv:0704.3084 [hep-th]].


\bibitem{Hirn:2005vk}
 J.~Hirn, N.~Rius and V.~Sanz,
  Phys.\ Rev.\  D {\bf 73}, 085005 (2006)
  [arXiv:hep-ph/0512240].

\bibitem{Erlich:2008gp}
  J.~Erlich and C.~Westenberger,
  Phys.\ Rev.\ D {\bf 79}, 066014 (2009)  
 [arXiv:0812.5105 [hep-ph]].

\bibitem{Polchinski:2000uf}
J.~Polchinski and M.~J.~Strassler,
  arXiv:hep-th/0003136.


\bibitem{Karch:2002xe}
 A.~Karch, E.~Katz and N.~Weiner,
  Phys.\ Rev.\ Lett.\  {\bf 90}, 091601 (2003)
  [arXiv:hep-th/0211107].


\bibitem{Hong:2004sa}
S.~Hong, S.~Yoon and M.~J.~Strassler,
  JHEP {\bf 0604}, 003 (2006)
  [arXiv:hep-th/0409118].

\bibitem{Witten:1998qj}
  E.~Witten,
  Adv.\ Theor.\ Math.\ Phys.\  {\bf 2}, 253 (1998)
  [arXiv:hep-th/9802150].

\bibitem{Gubser:1998bc}
 S.~S.~Gubser, I.~R.~Klebanov and A.~M.~Polyakov,
  Phys.\ Lett.\  B {\bf 428}, 105 (1998)
  [arXiv:hep-th/9802109].

\bibitem{Klebanov:1999tb}
  I.~R.~Klebanov and E.~Witten,
  Nucl.\ Phys.\  B {\bf 556}, 89 (1999)
  [arXiv:hep-th/9905104].

\bibitem{Kapusta}
T.~Gherghetta, J.~I.~Kapusta and T.~M.~Kelley,
  Phys.\ Rev.\  D {\bf 79}, 076003 (2009)
  [arXiv:0902.1998 [hep-ph]].

\bibitem{Becciolini:2009fu}
  D.~Becciolini, M.~Redi and A.~Wulzer,
  arXiv:0906.4562 [hep-ph].

\bibitem{Katz:2005ir}
  E.~Katz, A.~Lewandowski and M.~D.~Schwartz,
  Phys.\ Rev.\  D {\bf 74}, 086004 (2006)
  [arXiv:hep-ph/0510388].

\bibitem{KatzLatz}
E. Katz, plenary talk at Lattice 2008 conference.  Talk posted at
http://conferences.jlab.org/lattice2008/plenary.html


\bibitem{Kwee:2007nq}
H.~J.~Kwee and R.~F.~Lebed,
  Phys.\ Rev.\  D {\bf 77}, 115007 (2008)
  [arXiv:0712.1811 [hep-ph]].


\bibitem{Grigoryan:2007vg}
H.~R.~Grigoryan and A.~V.~Radyushkin,
  Phys.\ Lett.\  B {\bf 650}, 421 (2007)
  [arXiv:hep-ph/0703069].


\bibitem{Abidin:2008ku}
  Z.~Abidin and C.~E.~Carlson,
  Phys.\ Rev.\  D {\bf 77}, 095007 (2008)
  [arXiv:0801.3839 [hep-ph]].

\bibitem{Kovtun:2004de}
P.~Kovtun, D.~T.~Son and A.~O.~Starinets,
  Phys.\ Rev.\ Lett.\  {\bf 94}, 111601 (2005)
  [arXiv:hep-th/0405231].


\bibitem{Kovtun:2008kx}
 P.~Kovtun and A.~Ritz,
  Phys.\ Rev.\  D {\bf 78}, 066009 (2008)
  [arXiv:0806.0110 [hep-th]].


\bibitem{Breitenlohner:1982jf}
P.~Breitenlohner and D.~Z.~Freedman,
  Annals Phys.\  {\bf 144}, 249 (1982).



\bibitem{Karch:2006pv}
  A.~Karch, E.~Katz, D.~T.~Son and M.~A.~Stephanov,
  Phys.\ Rev.\  D {\bf 74}, 015005 (2006)
  [arXiv:hep-ph/0602229].



\bibitem{Cherman:2008eh}
  A.~Cherman, T.~D.~Cohen and E.~S.~Werbos,
  arXiv:0804.1096 [hep-ph].

\bibitem{Hofman:2008ar}
 D.~M.~Hofman and J.~Maldacena,
  JHEP {\bf 0805}, 012 (2008)
  [arXiv:0803.1467 [hep-th]].


\bibitem{Csaki:2008dt}
C.~Csaki, M.~Reece and J.~Terning,
  JHEP {\bf 0905}, 067 (2009)
  [arXiv:0811.3001 [hep-ph]].

\bibitem{pire}
B.~Pire, C.~Roiesnel, L.~Szymanowski and S.~Wallon,
  Phys.\ Lett.\  B {\bf 670}, 84 (2008)
  [arXiv:0805.4346 [hep-ph]].

\bibitem{Glozman}
L.~Y.~Glozman,
  arXiv:0903.3923 [hep-ph].


\end{thebibliography}
\end{document}